\documentclass[useAMS,usenatbib]{mn2e}

\usepackage{aas_macros,graphicx,multirow,rotate,amssymb}

\def\src{GRB\,061122}

\def\int {\emph{INTEGRAL}}

\title[The polarized GRB 061122]{The polarized Gamma-Ray Burst GRB 061122}
\author[D. G\"{o}tz et al.]{D. G\"{o}tz$^{1}$\thanks{E-mail:
diego.gotz@cea.fr}, S. Covino$^{2}$,  A. Fern\'andez-Soto$^{3,4}$, P. Laurent$^{5}$, \v Z. Bo\v snjak$^{1,6}$
\\
\smallskip\\
$^{1}$AIM (UMR 7158 CEA/DSM-CNRS-Universit\'e Paris Diderot) Irfu/Service d'Astrophysique, Saclay, F-91191 Gif-sur-Yvette Cedex, France\\
$^{2}$INAF -- Osservatorio Astronomico di Brera,  Via E. Bianchi 46, 23807 Merate (LC), Italy \\
$^{3}$Instituto de Fisica de Cantabria, CSIC-Universidad Cantabria, Avenida de los Castros s/n, 39005 Santander, Spain\\
$^{4}$Unidad Asociada Observatori Astronomic Universitat de Valencia - Instituto de Fisica de Cantabria, C/ Catedratico Jose Beltran 2, \\ 46980 Paterna, Spain\\
$^{5}$APC (UMR 7164 CEA/DSM/Irfu, Universit\'e Paris Diderot, CNRS/IN2P3, Observatoire de Paris) 10, rue Alice Domon et L\'eonie Duquet,\\ 75205 Paris Cedex 13, France\\
$^{6}$Department of Physics, University of Rijeka, 51 000 Rijeka, Croatia}
\begin{document}

\date{Accepted . Received ; in original form }

\pagerange{\pageref{firstpage}--\pageref{lastpage}} \pubyear{2013}

\maketitle

\label{firstpage}

\begin{abstract}

We report on the polarization measure, obtained with IBIS on board \int\, of the prompt emission of \src. Over an 8 s interval containing the brightest part of the Gamma-Ray Burst we put a lower limit on its polarization fraction of 60\% at 68\% c.l. and of 33\% at 90\% c.l. on the 250--800 keV energy range. 

We performed late time optical and near infra-red imaging observations of the GRB field using the Telescopio Nazionale Galileo (TNG), and the Canada-France-Hawaii Telescope (CFHT). Our multi-band ($ugrizYJHK$) photometry allowed us to identify the host galaxy of \src\ and to build its SED. Using a photometric redshift code we fitted these data, and derived the basic properties of the galaxy, including its type and redshift, that we could constrain to the interval [0.57, 2.10] at a 90\% c.l., with a best fit value of $z$=1.33.

The polarization measurement in different energy bands, together with the distance determination, allowed us to put the most stringent limit ($\xi \lesssim 3.4\times10^{-16}$) to date to a possible Lorentz Invariance Violation based on the vacuum birefringence effect, predicted by some quantum-gravity theories.

\end{abstract}

\begin{keywords}
gamma-rays burst: general -- gamma-rays burst: individual: GRB 061122 -- galaxies: photometry -- polarization -- gravitation
\end{keywords}

\section{Introduction}

Gamma-Ray Bursts (GRBs) are short lived transients (ms to hundreds of seconds) of soft $\gamma$-ray radiation that appear at random directions on the whole sky. Despite the recent progresses in the GRB field obtained mainly thanks to the \textit{Swift} and \textit{Fermi} satellites \citep[see e.g.][]{gehrels09,zhang11}, the nature of their prompt emission is still not clear. Nevertheless, thanks to the information obtained from their long-lived (hours to days) X-ray and optical counterparts, they have been proven to be of cosmological origin, with their redshifts, $z$, distributed in the range [$0.1,\sim9$], and some are firmly associated with Supernovae of type Ib/c.
These powerful explosions emit in a handful of seconds an amount of isotropic equivalent energy, $E_{\rm iso}$, that spans from 10$^{50}$ to 10$^{54}$ erg \citep[e.g.][]{amati07}, making them the most luminous events in the Universe, temporarily outshining all other sources. This huge amount of energy is nonetheless partially reduced, by accepting the hypothesis that GRBs are collimated sources \citep[e.g.][]{rhoads97}, and indeed the detection of some achromatic breaks in the light curves of GRB afterglows further supports the interpretation of GRBs as being produced in collimated jets, implying an energy reservoir of about 10$^{51}$ erg \citep{frail01,bloom03,ghirlanda12}.

However, the precise content of this jet, and especially its magnetization, as well as the details of the mechanism leading to the $\gamma$-ray emission are still not completely clear. Models range from unmagnetized fireballs where the observed emission could be produced by relativistic ($\Gamma \ga 100$) electrons accelerated in internal shocks propagating within the outflow \citep{rees94}, to pure electromagnetic outflows where the radiated energy comes from magnetic dissipation \citep{lyutikov06}. Intermediate cases with mildly magnetized outflows are of course possible \citep[e.g.][]{spruit01}.
Even in the case of an unmagnetized fireball, a local magnetic field in the emission region, generated by the shocks, is necessary if the dominant process is synchrotron radiation from relativistic electrons. 

Recently some measurements of polarization during the prompt emission of GRBs in the hundreds of keV energy range have been reported \citep[][]{kalemci07,mcglynn07,mcglynn09,gotz09,yonetoku11,yonetoku12}. These measurements can shed new light on the strength and scale of magnetic fields, as well as on the radiative mechanisms at work during the GRB prompt emission phase. 
Even if globally incoherent, in the case where the magnetic field is mainly transverse and locally highly ordered, i.e. has a local coherence scale which is larger than the typical size $\sim R / \Gamma$ of the visible part of the emitting region, the detected signal can still be highly polarized. On the other hand, in the case of a random field or an ordered magnetic field parallel to the expansion velocity, the polarization of the detected signal should vanish, except for the peculiar condition of a jet observed slightly off-axis \citep[e.g.][]{lazzati04}.

Polarization measures in cosmological sources are also a powerful tool to constrain Lorentz Invariance Violation (LIV), arising from the phenomenon of vacuum
birefringence as shown recently by \citet{laurent11a}, \citet{toma12}, and \citet{fan07}.
 
Here we present the polarization results on the prompt emission of \src, obtained with the IBIS telescope on board \int\ (section \ref{polar}), as well as the late time photometry of the GRB field observed with the Telescopio Nazionale Galileo (TNG) and the Canada-French-Hawaii Telescope (CFHT), that allowed us to identify the host galaxy and constrain its distance (section \ref{host}). Finally (section \ref{liv}) we present the limits we could derive on the possible LIV predicted by some quantum gravity theories \citep[like e.g. loop quantum gravity, see][]{gambini99}.

\section{Polarimetric Results}
\label{polar}

\subsection{\int\ Observations and Data Reduction}

\src\ has been detected by the INTEGRAL Burst Alert System \citep[IBAS;][]{ibas} on November 11$^{th}$ 2006, and localized to R.A. = $20^{h}15^{m}20.88^{s}$ Dec.= $+15^{\circ}30^{\prime}50.8^{\prime\prime}$, with an 90\% c.l. uncertainty of 2$^{\prime}$ \citep{mereghetti06}. With a peak flux of 31.7 ph cm$^{-2}$ s$^{-1}$, and a fluence of 2$\times$10$^{-5}$ erg cm$^{-2}$ (20--200 keV) it ranks second among the GRBs detected by \int, after GRB 041219A. It had a $T_{90}$ duration of 12 s \citep{vianello09}, and a moderately high peak energy of about 160--170 keV \citep{mcglynn09,golenetskii06,bosnjak13}.

IBIS \citep{ibis} is a coded mask telescope on board the \int\ satellite \citep{integral}. It is made by two pixellated detector layers, ISGRI \citep{isgri} working in the 15 keV--1 MeV energy range, and PICsIT \citep{picsit}, working in the 200 keV -- 10 MeV energy range. The two layers are superposed and permit to IBIS to be used as a Compton telescope by measuring the properties of the photons interacting in the two layers.  
Thanks to the polarization dependency of the differential cross section for Compton scattering -- linearly polarized photons scatter preferentially perpendicularly to the incident
polarization vector --, a Compton telescope can be used also as a polarimeter, and IBIS allowed us to detect polarization in three different objects, the Crab nebula \citep{forot08}, the black hole binary Cyg X--1 \citep{laurent11b}, and GRB 041219A \citep{gotz09}. In this work we adopt the same analysis technique as described in these references. 

Due to the nature of Compton scattering, one can expect an azimuthal distribution of the scattered photons on the telescope lower plane of the form

\begin{equation}
N(\phi)=S[1+a_{0}\cos 2(\phi-\phi_{0})],
\label{eq:azimuth}
\end{equation}

where $P.A. =  \phi_{0} - \pi /2 + n \pi$ is the polarization angle, and $\Pi= a_{0}/a_{100}$ the polarization fraction, where $a_{100}$ is the amplitude expected for a 100\% polarized source derived by Monte Carlo simulations of the instrument \citep[see e.g.][]{forot08}. 

To perform the polarization analysis, we derived the source flux as a function of $\phi$, and the scattered photons were then divided in 6 bins of 30$^{\circ}$ as a function of the azimuthal scattering angle. To improve the signal-to-noise ratio in each bin, we took advantage of the $\pi$-symmetry of the differential cross section, i.e. the first bin contains the photons with $0^{\circ}<\phi<30^{\circ}$ and $180^{\circ}<\phi<210^{\circ}$, etc.
The chance  coincidences (i.e. photons interacting in both detectors, but not related to a Compton event), have been subtracted from each detector image following the procedure described in \citet{forot08}. The derived detector images were then deconvolved to obtain sky images, where the flux of the source in each bin is measured by fitting the instrumental  PSF to the source peak, building a so-called polarigram of the source.

\subsection{Analysis and Results}

\label{sec:results}
\label{integral:polar}

\begin{figure}
\centering
\includegraphics[angle=0,width=8.5cm]{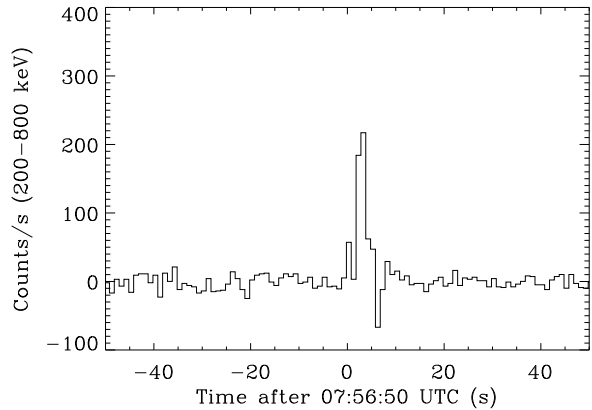}
\caption{IBIS Compton events background-subtracted light curve of \src. The data gap towards the end of the GRB is due to telemetry transmission limitations at satellite level.}
      \label{fig:lc}
\end{figure}

In Fig. \ref{fig:lc} we report the IBIS background-subtracted light curve of \src, derived using the Compton events in the 200--800 keV energy range.

\begin{figure}
\centering
\includegraphics[angle=90,width=8.5cm]{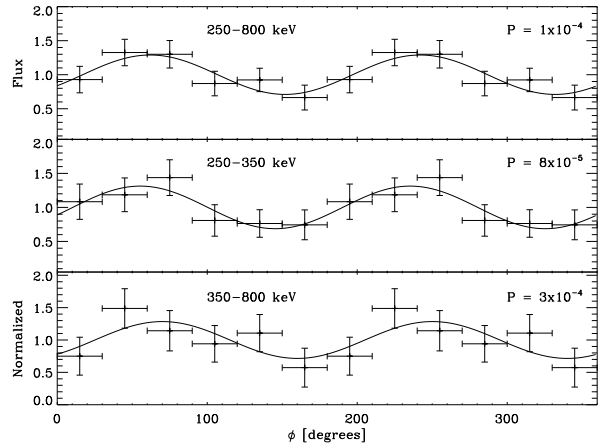}
\caption{Polarigrams of \src\ in different energy bands. The crosses
represent the data points (replicated once for clarity)
and the continuous line the fit done on the first 6 points using Eq. \ref{eq:azimuth}. 
The chance probability of a non-polarized ($<$1\%) signal
is reported in each panel.}
      \label{fig:polarigram}
\end{figure}

We performed the polarization analysis over different time intervals of the GRB. The best signal-to-noise ratio is obtained over the 07:56:50.0--07:56:58.0 U.T. time interval. In order to compute $a_{100}$, we derived the IBIS/ISGRI and SPI spectra over the same time interval with the technique developed by \citet{bosnjak13}, and fitted them jointly with XSPEC v. 12.3.0 \citep{xspec}. The data can be equally well be fitted with a Band model \citep{band93} or a power law with a high-energy cut-off. Since for the former the $\beta$ parameter is not constrained we use the latter. The best fit photon index is $\alpha$=-1.15$\pm$0.04, and the cut-off energy $E_{c}$=221$\pm$20 keV (both errors are given at 90\% c.l.), corresponding to a peak energy of 188 keV, slightly higher than the one of the average spectrum. Given these spectral parameters, $a_{100}$ has been computed through Monte Carlo simulations, and turns out to be independent from the energy band, being 0.29$\pm$0.02 in the 250--800 keV energy band, 0.30$\pm$0.03 in the 250--350 keV energy band, and 0.29$\pm$0.03 in the 350--800 keV energy band. We built the corresponding polarigrams in three energy bands (250--800, 250--350, 350--800 keV), and we fitted them with Eq. \ref{eq:azimuth} using a least squares technique to derive $a_{0}$ and $\phi_{0}$, see Fig. \ref{fig:polarigram}. Confidence intervals on $a_{0}$ and $\phi_{0}$ were, on the other hand, not derived from the fit, since the two variables are not independent. They were derived from the probability density distribution of measuring $a$ and $\phi$ from $N$ independent data points over a $\pi$ period, based on Gaussian distributions for the orthogonal Stokes components (see Eq. 2 in \citealt{forot08}). 

\begin{figure}
\includegraphics[angle=90,width=8.6cm]{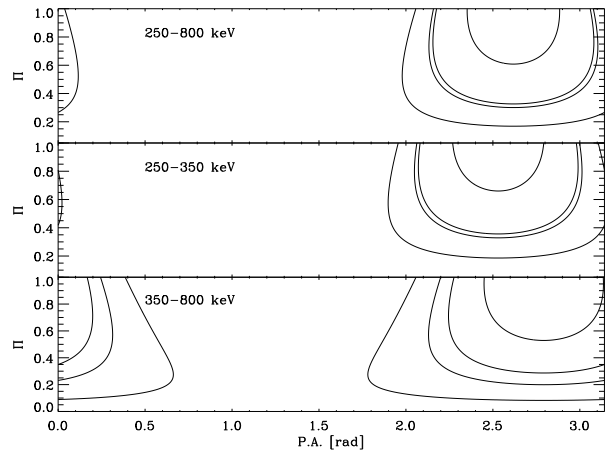}
\caption{The 68\%, 90\% 95\%, and 99\% (top to bottom in each panel) confidence contours for the $\Pi$ and $P.A.$ parameters for three energy ranges.}
      \label{fig:errors}
\end{figure}

Over the selected time interval we measure a high polarization level in the 250--800 keV energy band, deriving a 68\% c.l. lower limit to the polarization fraction ($\Pi$) of 60\%
and the polarization angle is 150$\pm$15$^{\circ}$. The corresponding polarigram is shown in the upper panel of Fig. \ref{fig:polarigram}. The 68\%, 90\%, 95\%, and 99\% confidence regions for the two parameters are shown in Fig. \ref{fig:errors}, where one can see that the 90\% and 99\% c.l. lower limit to $\Pi$ are 33\% and 16\%, respectively. So despite the fact that the statistics is not high enough to fully constrain the polarization fraction, we can exclude that our signal is due to an un-polarized source ($\Pi<$1\%) at a probability level $P$ of 10$^{-4}$ (also shown in Fig. \ref{fig:polarigram}).
The same analysis has been performed for the 250--350 and 350--800 keV energy bands, and the results are plotted also in Figs. \ref{fig:polarigram} and \ref{fig:errors}, and reported in Table \ref{tab:pola}.

\begin{table}
\centering
\caption{Polarization measurements of \src.} 
\label{tab:pola}
\begin{tabular}{ccccc}
\hline
Energy band & $\Pi$ (\%)   & P.A. ($^{\circ}$) & $\Pi$ (\%) & P.A. ($^{\circ}$)\\ 
(keV) & (68\% c.l.) & (68\% c.l.) & (90\% c.l.) & (90\% c.l.)\\ 
\hline
250--800 & $>$60 & 150$\pm$15 & $>$33 & 150$\pm$20 \\
250--350 & $>$65 & 145$\pm$15 & $>$35 & 145$\pm$27 \\
350--800 &$>$52 & 160$\pm$20 & $>$20 & 160$\pm$38 \\
\hline
\end{tabular}
\end{table}

\section{Host Galaxy Identification}
\label{host}
\subsection{TNG Observations}

The \src\ field was observed with Dolores and NICS at the 3.5 m TNG in La Palma (Canary Islands,  program ID: AOT24/TAC\_12) in queued observing mode. The Dolores data have been collected between August 1$^{st}$ and 5$^{th}$ 2011 in the $ r^{\prime} $,$ i^{\prime} $, and $ z^{\prime} $ filters, while the NICS data have been obtained between August 12$^{th}$ and 15$^{th}$ 2011 in the $ J $, and $ K_{s} $ filters. During the Dolores observations the seeing was in the 0.5--1.0$^{\prime\prime}$ interval, while during the NICS ones it ranged between 0.7$^{\prime\prime}$ and 1.0$^{\prime\prime}$. Data for the TNG run (and the CHFT run) are reported in Table \ref{tab:expo}. 

TNG data were reduced by means of a custom pipeline\footnote{http://pypi.python.org/pypi/SRPAstro/} implementing standard procedures. Data were bias/dark subtracted and flat-field correction was applied. Individual frames were shifted to a common reference and astrometry was derived on the final averaged images using the 2MASS catalogue\footnote{http://www.ipac.caltech.edu/2mass/} as a reference. Aperture photometry was performed by means of the GAIA\footnote{http://star-www.dur.ac.uk/$\sim$pdraper/gaia/gaia.html} tools and calibration was derived with suitable isolated non-saturated stars in the field using the photometry from the 2MASS catalogue in the NIR. For the optical data we relied on the calibration carried out for the CFHT frames (see Sect.\,\ref{sec:chft}).

\subsection{CFHT Observations}
\label{sec:chft}
The \src\ field was observed with MegaCam and WIRCam at the 3.6 m CFHT in Mauna Kea (Hawaii, program ID: 11BF020) in queued service observation mode. The MegaCam data were collected between September 6$^{th}$ and October 31$^{st}$ in the $ u ^{*}$, $ g^{\prime} $,$ r^{\prime} $,$ i^{\prime} $, and $ z^{\prime} $ filters, while the WIRCam data were obtained between September 12$^{th}$ and October 12$^{th}$ 2011 in the $ Y $, $ J $, $ H $, and $ K_{s} $ filters. During the MegaCam observations the seeing was in the 0.8--1.2$^{\prime\prime}$ interval, while during the WIRCam ones it ranged between 0.5$^{\prime\prime}$ and 0.8$^{\prime\prime}$. 
Due to bad weather conditions, the MegaCam observation program could not be completed resulting in the reduced exposure times reported in Table \ref{tab:expo}. In the same table we report also the exposure for the NIR filters obtained with WIRCam.

As described in \citet{gotz11}, the CFHT data were reduced using the $TERAPIX$\footnote{http://terapix.iap.fr} pipeline. The latter retrieves
pre-processed data from the CFHT database, combines the mask and gain-maps of the individual images, -- obtaining the weight-maps, that are needed to perform image mosaicking -- subtracts bias/dark images, and produces flat-field images. Astrometry is then performed on individual images using a reference catalogue (USNO\footnote{http://www.usno.navy.mil/USNO/astrometry/optical-IR-prod}, 2MASS) and a pattern matching algorithm. Then, overlapping detections are identified among the individual images, and a global astrometric solution is computed. Photometry is then performed in a similar way, harmonizing the zero-points from the different pointings to account for different atmospheric extinctions and non-photometric observing conditions. Images are finally co-added, using the astrometric and photometric calibrations, and the weight-maps described above.  

As for the TNG data, the GAIA package was used to perform the aperture photometry, and the final photometry, reported in Table \ref{tab:expo}, was derived by a weighted mean of the TNG and CFHT results when both are available.

\subsection{GTC Observations}

We obtained observing time on the 10.4 m Gran Telescopio Canarias (GTC) in the Observatorio del Roque de los Muchachos (Canary Islands, Spain) to perform spectroscopy of the GRB host galaxy candidate. Our initial program (ID 44-GTC17/11B) was accepted but could not be carried out, and our second program (37-GTC8/12B) was successful. The observations were performed in queue mode on the night of August 20$^{th}$ 2012, under dark, clear conditions, with seeing approximately 1.0$\arcsec$. 

GTC provided us with long-slit spectroscopic observations of the brightest candidate compatible with our early finding chart of the GRB field, obtained at TNG, where object 2 was the only detected source close to the XRT error box (see Section \ref{sec:identification}). We used the R300R grating ($R \approx 300$) and integrated for 4200 seconds on target. Standard calibration images and procedures were acquired and followed for wavelength calibration, and an approximate flux calibration was obtained using the spectrophotometric standard star GD190.

\subsection{Photometry, Spectroscopy, and Identification}

\subsubsection{Imaging}
\label{sec:identification}

\begin{figure}
\centering
\includegraphics[angle=0,width=8.5cm]{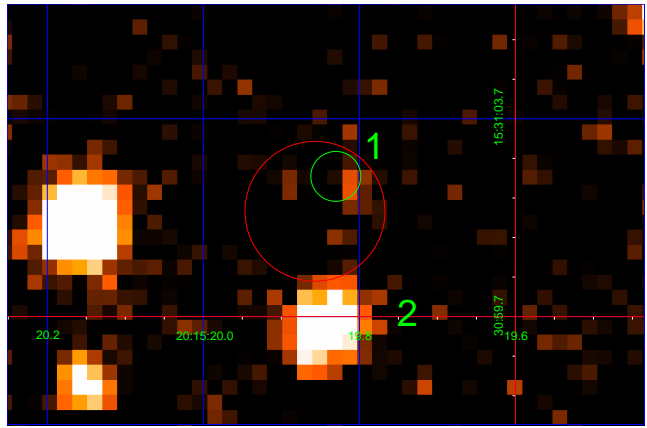}
\caption{CFHT/WIRCam image of the \src field obtained in the $J$ filter. The red circle represents the Swift/XRT error box, while the smaller green circle represents the optical afterglow error box. Object 1 has been identified as the GRB host galaxy candidate.} 
      \label{fig:wircamj}
\end{figure}

We have compared the most precise error boxes provided in the literature for \src, with our images. In particular we used the Swift/XRT position, derived using 26 ks of data \citep{evans10}, provided by the Swift Science Data Centre (http://www.swift.ac.uk), corresponding to R.A. =  $20^{h}15^{m}19.87^{s}$ and Dec. = $+15^{\circ}31^{\prime}01.8^{\prime\prime}$ with an uncertainty of 1.4$^{\prime\prime}$, and the optical afterglow position published by \citet{halpern06}, obtained in the $R$ band, and corresponding to R.A. =  $20^{h}15^{m}19.84^{s}$ and Dec. = $+15^{\circ}31^{\prime}02.5^{\prime\prime}$ with an uncertainty of 0.5$^{\prime\prime}$. 
As can be seen from our CFHT/WIRCam $J$ band image (Fig. \ref{fig:wircamj}) there is only one bright object (labelled ``2'') that is marginally compatible with the XRT error box (drawn in red), while there is a faint object (labelled ``1'') that is compatible with both error boxes. We therefore identify the \src\ host galaxy candidate with object 1. The latter is located at R.A. = $20^{h}15^{m}19.81^{s}$ Dec. = $+15^{\circ}31^{\prime}02.5^{\prime\prime}$, and its measured magnitudes based on both datasets are reported in Table \ref{tab:expo}.


\begin{table}
\centering
\caption{Exposure times and measured magnitudes (1 $\sigma$ c.l.) for the host galaxy of GRB\,061122. Galactic extinction along the line of sight has not been subtracted from the data.} 
\label{tab:expo}
\begin{tabular}{ccccc}
\hline
Filter & CFHT Exp. & TNG Exp.  & Mag. & Mag.\\ 
(System) & Time (ks) & Time (ks) & Object 1 & Object 2\\ 
\hline
$ u^{*} (AB)$& 4.5&&$>$24.5 (3$\sigma$)&$>$24.75 (3$\sigma$)\\
$ g^{\prime}(AB) $& 1.5 &&$>$24.25 (3$\sigma$)&23.60$\pm$0.12\\
$ r^{\prime} (AB)$& 1.2 & 1.9 &$>$24.0 (3$\sigma$)&23.15$\pm$0.10\\
$ i^{\prime} (AB) $& 1.1& 1.8 &$>$23.5 (3$\sigma$)&22.23$\pm$0.08\\
$ z^{\prime} (AB)$& 0.5& 1.8 &$>$23.0 (3$\sigma$)&21.74$\pm$0.10\\
$ Y (Vega) $& 7.6 &&22.66$\pm$0.25&20.64$\pm$0.07\\
$ J (Vega)$& 7.6& 2.6 &22.31$\pm$0.27&20.16$\pm$0.06\\
$ H (Vega)$& 4.9&&21.52$\pm$0.33&19.65$\pm$0.09\\
$ K_{s} (Vega) $&5.7& 5.3 &20.63$\pm$0.22&19.37$\pm$0.10\\
\hline
\end{tabular}
\end{table}


The source appears to be point-like in all frames where it is detected, although its faintness would have probably prevented any extension be measured. It is detected only in the NIR, and this might be consistent with a high redshift source or a dusty sight-line. However, the not negligible Galactic extinction along the line of sight,  $E_{B-V} = 0.16$ \citep{schlafly11}, makes optical detection, in particular in the bluer filters, even more demanding. 

\subsubsection{GTC Spectroscopy}

\begin{figure}
\includegraphics[angle=-90,width=8.6cm]{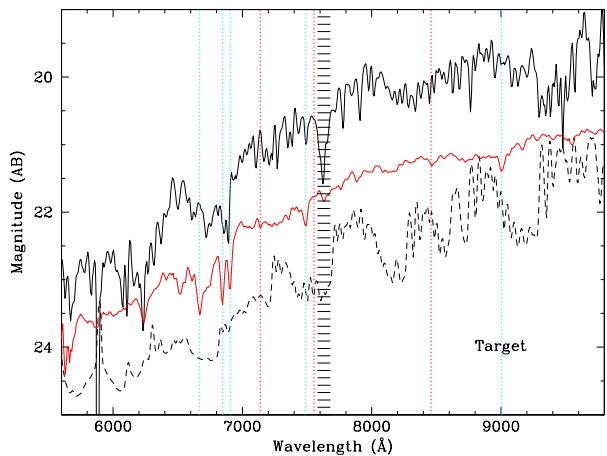}
\caption{GTC spectrum of object 2 and its associated 1$\sigma$ uncertainty are plotted as the continuous and dashed lines. The red line shows for comparison the template spectrum of an elliptical galaxy at redshift $z=0.74$, with the vertical lines marking the position of the main emission and absorption features. The shaded area at $\lambda \approx 7600$\AA\ corresponds to the main atmospheric O$_2$ line.}
      \label{fig:gtcspec}
\end{figure}

When this project started, prior to our CFHT/WIRCam observations, the object labelled as ``object 2'' (AB$(R)\approx 23.2$) in Fig. \ref{fig:wircamj} was the only detected candidate which was barely compatible with the Swift/XRT observations. Hence, this is the object for which we performed the spectroscopic observations in order to measure its redshift. As was mentioned above we obtained time on the GTC telescope, and in Fig. \ref{fig:gtcspec} we show its spectrum, which does not feature prominent emission or absorption lines. Based on the presence of a possible 4000\AA\ break in the photometric data, and associated absorption lines, we adopt $z=0.74$ as its redshift. The spectrum corresponds to that of a galaxy with an evolved stellar population, strengthening object 1 as being \src\ host galaxy candidate, see below.


\subsubsection{SED modelling and distance determination}

We have combined all our photometric observations from TNG and CFHT to produce a wide coverage spectral energy distribution ($ugrizYJHK$) for the putative host of \src, see Table \ref{tab:expo}. We have corrected it for the observed galactic extinction ($E_{B-V} = 0.16$), and using these data and the photometric redshift code by \citet{fernandez99}, we have estimated the redshift and basic properties of our host galaxy candidate, object 1. 

\begin{figure}
\includegraphics[angle=-90,width=8.6cm]{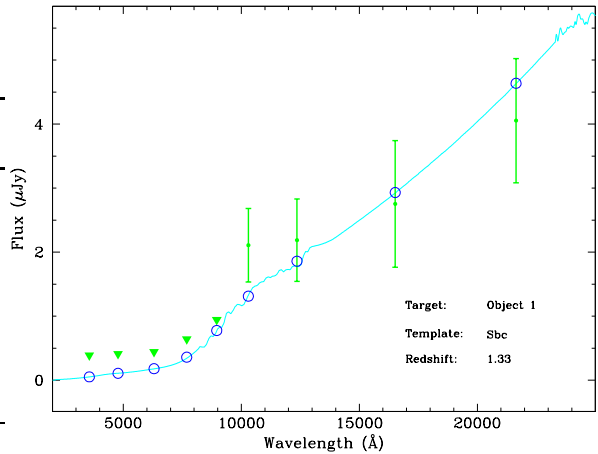}
\caption{The plot shows our TNG and CFHT photometry for the host galaxy candidate as filled green circles with error bars, and as green arrows in the case of our 1 $\sigma$ upper limits.  The best-fit spectrum obtained for an Sb/c galaxy template at $z=1.33$ is shown as a cyan line, and the expected photometry for such a model is indicated by the empty blue circles.}
      \label{fig:sed}
\end{figure}

As shown in Table \ref{tab:expo}, object 1 is not detected in our $ugriz$ images, and it is marginally seen in all the observed NIR bands ($YJHK$, in all of them between 3 and 5 sigma significance). At face value this may point to a very high-redshift solution ($z \approx 7$), but such a galaxy would hardly be detectable in our data. Excluding this very high redshift solution, the best-fit candidate is represented by a galaxy with an old stellar population template with redshift $z$ $\epsilon$ [0.38, 1.96].\footnote{All of the redshift intervals quoted are given at the 90\% confidence level, and include a systematic error component calculated as in \citet{fernandez02}.}  However, old stellar population galaxies are less likely to host long GRBs, which are rather associated with galaxies with on-going star formation \citep[see e.g.][]{savaglio09}. Indeed, the second best fit candidate for our photometry is represented by an Sb/c template at a slightly higher redshift $z$ $\epsilon$ [0.57, 2.10]. The best-fit solution in this case, with $z=1.33$, is shown in Fig. \ref{fig:sed}, and implies an isotropically equivalent emitted energy for the GRB of $E_{iso}\sim$3$\times$10$^{52}$ erg (1 keV--10 MeV).

\section{LIV limits}
\label{liv}

On general grounds one expects that the two fundamental theories of contemporary physics, the theory of General Relativity and the quantum theory in the form of the Standard Model of particle physics, can be unified at the Planck energy scale. This unification requires to quantize gravity, which leads to very fundamental difficulties: one of these is the possibility of Lorentz Invariance Violation (LIV) \citep[e.g.][]{jacobson06,liberati09,mattingly05}

A possible experimental test of LIV is testing the helicity dependence of the propagation velocity of photons \citep[see e.g.][and references therein]{laurent11a}. The light dispersion relation is given in this case by

\begin{equation}
\omega^{2}=k^{2}\pm\frac{2\xi k^{3}}{M_{Pl}}\equiv\omega^{2}_{\pm}
\label{eq:dispersion1}
\end{equation}

where $E=\hbar\omega$, $p=\hbar k$, $M_{Pl}$ is the Planck Mass, and the sign of the cubic term is determined by the chirality (or circular polarization) of the photons, which leads to a rotation of the polarization during the propagation
of linearly polarized photons. This effect is known as vacuum birefringence.

Equation \ref{eq:dispersion1} can be approximated as follows

\begin{equation}
\omega_{\pm}=\vert p \vert \sqrt{1\pm\frac{2\xi k}{M_{Pl}}}\approx\vert k\vert(1\pm\frac{\xi k }{M_{Pl}})
\label{eq:dispersion2}
\end{equation} 

where $\xi$ gives the order of magnitude of the effect. In other words, if a polarized signal is measured from a distant source, some quantum-gravity theories \citep[e.g.][]{myers03} predict that the polarization plane should rotate by a quantity $\Delta\theta$ while the electromagnetic wave propagates through space, and this as a function of the energy of the photons. This is illustrated in Eq. \ref{eq:rotation}, where $d$ is the distance of the source:

\begin{equation}
\Delta\theta(p)=\frac{\omega_{+}(k)-\omega_{-}(k)}{2}\ d\approx\xi\frac{k^{2}d}{2M_{Pl}}
\label{eq:rotation}
\end{equation}

This implies that a polarized signal produced by a given source would vanish if observed on a broad energy band, since the differential rotation acting on the polarization angle as a function of energy would in the end add opposite oriented polarization vectors.
But being this effect very tiny, since it is inversely proportional to the the Planck Mass ($M_{Pl}\sim$2.4$\times$10$^{18}$ GeV), the observed source needs to be at cosmological distances. So, the simple fact to detect the polarization signal from a distant source, can put a limit to such a possible violation. This experiment has been performed recently by \citet{laurent11a} and \citet{toma12} making use of GRBs. 
Indeed, since GRBs are cosmological, their polarization measurements are highly suited to measure and improve upon these limits.
\citet{laurent11a}, taking advantage from the polarization measurements obtained with IBIS on GRB 041219A in different energy bands (200--250 keV, 250--325 keV), and from the measure of distance of the source (z$>$0.02 at 90\% c.l., equivalent to a luminosity distance 85 Mpc) were able to set the most stringent limit to date to a possible LIV effect: $\xi <$1.1$\times$10$^{-14}$. We note that, although \citet{toma12} claim to have derived a more stringent limit ($\xi <$8$\times$10$^{-16}$), their measure does not rely on a real measure of the distance of the GRBs they analyse, but they use a distance estimate based on an empirical spectral-luminosity relation \citep{yonetoku10}, whose selection effects, physical interpretation, and absolute calibration are not yet completely understood.

By taking the 90\% c.l. lower limit on the distance of the host of \src\ we derived through our multi-band SED modelling, $z$ = 0.54, corresponding to a luminosity distance of 3.309 Gpc ($\Omega_{m}$ = 0.27, $\Omega_{\lambda}$ = 0.73, and H$_{0}$ = 71 km/s/Mpc), and if we set $\Delta\theta(k)$ = 80$^{\circ}$ (see Tab. \ref{tab:pola}), we obtain

\begin{equation}
\xi < \frac{2 M_{Pl}\Delta\theta(k)}{(k_{2}^{2}-k_{1}^{2})\ d}\approx 3.4 \times 10^{-16}.
\label{eq:xi}
\end{equation}

\section{Discussion and Conclusions}

We measured linear polarization in the $\gamma$-ray energy band (250--800 keV) during the brightest part of the prompt emission of GRB 061122. We were able to put a lower limit on the polarization level of 33\% (90\% c.l.), and exclude an un-polarized signal at a probability level of 10$^{-4}$. This measure, follows some recent reports of detections of high (and variable) polarization levels in the prompt emission of a few other GRBs: 041219A by \citet{gotz09},  100826A, 110301A and  110721A by \citet{yonetoku11,yonetoku12}. Although all these measures, taken individually, have not a very high significance ($\gtrsim$3 $\sigma$), they indicate that GRBs may indeed be emitters of polarized radiation. In addition, for the \int\ bursts GRB 041219A, and GRB 061122 there are independent and compatible measurements obtained by SPI and IBIS: for 061122 \citet{mcglynn09}, using SPI data, derived a measure (compatible with the one presented here) of $\Pi$=29$_{+25}^{-26}$\% (1 $\sigma$ c.l., 100 keV--1 MeV) for the polarization level, but they could not constrain further this results due to lack of statistics; for bright GRB 041219A \citet{mcglynn07} and \citet{gotz09} measured independently on a short interval during the GRB peak emission $\Pi$=68$\pm$29\% (100 keV--1 MeV), and 65$\pm$26\% (250 keV--800 keV), respectively. These are remarkably similar results, while the apparent discrepancy, pointed out by \citet{yonetoku11,yonetoku12} on the measure of the polarization on a longer time interval, where \citet{gotz09} put an upper limit on the polarization level of $\Pi<$4\%, while \citet{mcglynn07} detects a signal of 26$\pm$20\%, can be easily reconciled by the fact that both measures provide errors at a 1 $\sigma$ c.l. In addition, the measurement on the longer time interval containing the peak is statistically largely dominated by the peak itself in the SPI data (the measured P.A. is actually the same for both intervals), while this is not the case in the IBIS data: due to the larger effective area with respect to SPI, IBIS records a much larger number of counts for a given source. But this has the drawback of losing some of the counts during transmission because of telemetry limitations at satellite level. The consequence for GRB 041219A was that in the IBIS data the bright peak (i.e. the short interval) had about the same number of counts than its wings, and hence the statistical coverage of the burst was more homogeneous over its different parts, allowing us, on one hand, to measure the variations of the P.A., but implying, on the other hand, the net effect that the different portions of the GRB, with different P.A., tend to wash out the polarization signal on a longer time interval.

As discussed in the works mentioned above, the measure of a high level of linear polarization, as well as its variability, point towards an interpretation where synchrotron radiation is emitted from shock accelerated electrons in a relativistic jet with a magnetic field transverse to the jet expansion. The coherence of the magnetic field geometry does not need to hold over the entire jet, but only over a small portion of it, since, due to relativistic effects, the observer can see only a region of the jet whose angular size is comparable to 1/$\Gamma$, $\Gamma$ being the Lorentz factor of the relativistic outflow. If the radiating electrons are accelerated in internal shocks \citep{rees94,kobayashi97,daigne98}, then the Lorentz factor is necessarily varying in the outflow, which can explain the variability of the polarization from one pulse to the other \citep{granot03a,granot03b,nakar03}, as observed in GRB 041291A and GRB 100826A. This is not the only possible interpretation, and other models that predict locally coherent magnetic fields, like fragmented fireballs (shotguns, cannonballs, sub-jets) can produce highly polarized emission, with a variable polarization angle. In this frame the fragments are responsible for the single pulses and have different Lorentz factors, opening angles and magnetic domains \citep[e.g.][]{lazzati09}. In addition, different emission mechanisms can not be completely excluded at this time, implying random magnetic fields and peculiar observing conditions, like e.g. inverse Compton scattering \citep{lazzati04}. Only when more data with a higher accuracy become available, some model will be preferred on the basis of statistical arguments. On the other hand, to be able to quantitatively compare the data with the theory, more accurate models are also needed.

Thanks to our late time imaging of the field of \src\ obtained with the TNG and the CFHT, we were able to identify the host galaxy of GRB 061122, and to constrain through multi-band optical and NIR SED modelling, its type, and its distance to the redshift interval [0.57, 2.10]. The latter together with the polarization measure obtained with IBIS, allowed us to derive the deepest and most reliable limit available to date ($\xi <$3.4$\times$10$^{-16}$) on the possibility of Lorentz Invariance Violation, measured through the vacuum birefringence effect on a cosmological source.

\section*{Acknowledgements}

Based on observations with INTEGRAL, an ESA project with instruments and
science data centre funded by ESA member states (especially the PI
countries: Denmark, France, Germany, Italy, Switzerland, Spain), Czech
Republic and Poland, and with the participation of Russia and the USA,
on observations obtained with WIRCam, a joint project of CFHT, Taiwan, Korea, Canada, France, and with MegaCam, a joint project of CFHT and CEA/Irfu,
at the Canada-France-Hawaii Telescope (CFHT) which is operated by the National Research Council (NRC) of Canada, the Institute National des Sciences de l'Univers of the Centre National de la Recherche Scientifique of France, and the University of Hawaii, and on observations made with the Gran Telescopio Canarias (GTC), installed in the Spanish Observatorio del Roque de los Muchachos of the Instituto de Astrof\'{\i}sica de Canarias, on the island of La Palma.
ISGRI has been realized and maintained in flight by CEA-Saclay/Irfu with
the support of CNES. 
The authors are grateful to the TERAPIX team ({\tt http://terapix.iap.fr/}) for providing the CHFT data reduction. 
A.F.S. acknowledges support from the Spanish MICYNN projects AYA2010-22111-C03-02 and Consolider-Ingenio 2007-32022, and from the Generalitat Valenciana project Prometeo 2008/132.

\bibliographystyle{mn2e}
\bibliography{biblio}

\begin{thebibliography}{34}
\expandafter\ifx\csname natexlab\endcsname\relax\def\natexlab#1{#1}\fi



\bibitem[{{Amati} {} (2007){Amati}, L.}]{amati07}
{Amati} L., 2007, \mnras, 372, 233

%
\bibitem[{{Arnaud}{}(1996){Arnaud}, K.~A.}]{xspec}
{Arnaud} K.~A., 1996, Astronomical Society of the Pacific Conference Series, 101, 17

\bibitem[{{Band} {et al.}(1993){Band}, D. and {Matteson}, J. and {Ford}, L. and {Schaefer}, B. and 
	{Palmer}, D. and {Teegarden}, B. and {Cline}, T. and {Briggs}, M. and 
	{Paciesas}, W. and {Pendleton}, G. and {Fishman}, G. and {Kouveliotou}, C. and 
	{Meegan}, C. and {Wilson}, R. and {Lestrade}, P.}]{band93}
{Band} D. et al., 1993, \apj, 413, 281

\bibitem[{{Bloom} {et al.}(2003){Bloom}, J.~S. and {Frail}, D.~A. and {Kulkarni}, S.~R.}]{bloom03}
{Bloom} J.~S., {Frail} D.~A., {Kulkarni} S.~R., 2003, \apj, 594, 674
%
%


\bibitem[{{Bo{\v s}njak} {et al.} (2013)}]{bosnjak13}
{Bo{\v s}njak} {\v Z}. et al., 2013, submitted


%
 

%




%
\bibitem[{{Daigne} {\& Mochkovitch}(1998){Daigne}, F. and {Mochkovitch}, R.}]{daigne98}
Daigne F., Mochkovitch R., 1998, MNRAS, 296, 275


\bibitem[{{Di Cocco} {et~al.}(2003){Di Cocco}, G. and {Caroli}, E. and {Celesti}, E. and {Foschini}, L. and 
	{Gianotti}, F. and {Labanti}, C. and {Malaguti}, G. and {Mauri}, A. and 
	{Rossi}, E. and {Schiavone}, F. and {Spizzichino}, A. and {Stephen}, J.~B. and 
	{Traci}, A. and {Trifoglio}, M.}]{picsit}
{Di Cocco} G. {et~al.}, 2003, \aap, 411, L189

%

\bibitem[{{Evans} {et~al.} (2010)}]{evans10}
Evans P.~A. et al., 2010, \aap, 519, A102

\bibitem[{{Fan} {et~al.} (2007)}]{fan07}
{Fan} Y.-Z., {Wei} D.-M., {Xu} D., 2007, \mnras, 376, 1857

\bibitem[{{Fernandez-Soto} {et~al.} (1999){Fern{\'a}ndez-Soto}, A. and {Lanzetta}, K.~M. and {Yahil}, A.}]{fernandez99}
{Fern{\'a}ndez-Soto} A., {Lanzetta} K.~M., {Yahil} A., 1999, \apj, 513, 34

\bibitem[{{Fernandez-Soto} {et~al.} (2002)}]{fernandez02}
{Fern{\'a}ndez-Soto} A., {Lanzetta} K.M, {Chen} H.-W., {Levine} B., {Yahata} N., 2002, \mnras, 330, 889


\bibitem[{Forot} {et al.}(2008)]{forot08}
Forot M., Laurent P., Grenier I.A., Gouiff\`es C., Lebrun F., 2008, \apj, 688, L29


\bibitem[{{Frail} {et~al.} (2001){Frail}, D.~A. and {Kulkarni}, S.~R. and {Sari}, R. and {Djorgovski}, S.~G. and 
	{Bloom}, J.~S. and {Galama}, T.~J. and {Reichart}, D.~E. and 
	{Berger}, E. and {Harrison}, F.~A. and {Price}, P.~A. and {Yost}, S.~A. and 
	{Diercks}, A. and {Goodrich}, R.~W. and {Chaffee}, F.}]{frail01}
{Frail} D.~A. et~al., 2001, \apjl, 562, L55



\bibitem[{{Gambini} {\& Pullin}(1999)}]{gambini99}
Gambini R., Pullin J., 1999, Phys. Rev. D, 59, 124021

\bibitem[{{Gehrels} {et~al.}(2009){Gehrels}, {Ramirez-Ruiz}, {Fox}}]{gehrels09}
{Gehrels} N., {Ramirez-Ruiz} E., {Fox} D.~B., 2009, \araa, 47, 567



\bibitem[{{Ghirlanda} {et~al.}(2012)}]{ghirlanda12}{Ghirlanda} G., {Nava} L., {Ghisellini} G., {Celotti} A., {Burlon} D., {Covino} S., {Melandri} A., 2012, \mnras, 420, 483


%

\bibitem[{{Golenetskii} {et~al.}(2006)}]{golenetskii06}
Golenetskii S., Aptekar R., Mazets E., Pal'shin V., Frederiks D., Cline T., 2006, GCN, 5841


\bibitem[{{G\"otz} {et~al.}(2009){Gotz}, {Laurent}, {Lebrun}, {Daigne},	{Bosnjak}}]{gotz09}
{G\"otz} D., {Laurent} P., {Lebrun} F., {Daigne} F., {Bo{\v s}njak} {\v Z}., 2009, \apj, 695, L208

\bibitem[{{G\"otz} {et~al.}(2011)}]{gotz11}
{G{\"o}tz} D., {Covino} S., {Hasco{\"e}t} R., {Fernandez-Soto} A., 
	{Daigne} F., {Mochkovitch} R., {Esposito}, P. 2011, \mnras, 413, 2173
	
\bibitem[{{Granot} {}(2003)}]{granot03a} Granot J., 2003, \apjl, 596, L17

\bibitem[{{Granot} {\& K\"{o}nigl}(2003)}]{granot03b} Granot J., K\"{o}nigl A., 2003, \apjl, 594, L83



	
\bibitem[{{Halpern} {\& Armstrong}(2006)}]{halpern06}
Halpern J., Armstrong E., 2006, GCN, 5849



%



\bibitem[{{Jacobson} {et~al.}(2006)}]{jacobson06}
Jacobson T., Liberati S., Mattingly D., 2006, Annals of Physics, 321, 150

\bibitem[{{Kobayashi} {et al.}(1997)}]{kobayashi97} Kobayashi S., Piran T., Sari R., 1997, \apj, 490, 92


\bibitem[{{Kalemci} {et~al.}(2007)}]{kalemci07}
{Kalemci} E., {Boggs} S.~E., {Kouveliotou} C., {Finger} M., 
	{Baring} M.~G., 2007, \apjs, 169, 75 



\bibitem[{{Laurent} {et al.}(2011a)}]{laurent11a}
{Laurent} P., {G{\"o}tz} D., {Bin{\'e}truy} P., {Covino} S., {Fernandez-Soto} A., 2011, \prd, 83, 12, 121301
	
\bibitem[{{Laurent} {et al.}(2011b)}]{laurent11b}
{Laurent} P., {Rodriguez} J., {Wilms} J., {Cadolle Bel} M., 
	{Pottschmidt} K., {Grinberg} V., 2011, Science, 332, 438 
	

\bibitem[{{Lazzati} {et al.}(2004)}]{lazzati04} Lazzati D., Rossi E., Ghisellini G., Rees M.J., 2004, \mnras, 347, L1


\bibitem[{{Lazzati} {\& Begelman}(2009)}]{lazzati09} Lazzati D., Begelman M.C., 2009, \apj, 700, L141


\bibitem[{{Lebrun} {et~al.}(2003){Lebrun}, {Leray}, {Lavocat}, {Cr{\'e}tolle},
  {Arqu{\`e}s}, {Blondel}, {Bonnin}, {Bou{\`e}re}, {Cara}, {Chaleil}, {Daly},
  {Desages}, {Dzitko}, {Horeau}, {Laurent}, {Limousin}, {Mathy}, {Mauguen},
  {Meignier}, {Molini{\'e}}, {Poindron}, {Rouger}, {Sauvageon}, \&
  {Tourrette}}]{isgri}
{Lebrun} F. {et~al.}, 2003, \aap, 411, L141

\bibitem[{{Liberati} {\& Maccione}(2009)}]{liberati09}
Liberati S., Maccione L., 2009, Annual Review of Nuclear and Particle Science, 59, 245 

\bibitem[{{Lyutikov}{}(2006)}]{lyutikov06} Lyutikov M., 2006, New Journal of Physics, 8, 199


\bibitem[{{Mattingly} {}(2005)}]{mattingly05}
Mattingly D., 2005, Living Reviews in Relativity, 8, 5

\bibitem[{{McBreen} {et~al.}(2006){McBreen}, S. and {Hanlon}, L. and {McGlynn}, S. and {McBreen}, B. and 
	{Foley}, S. and {Preece}, R. and {von Kienlin}, A. and {Williams}, O.~R.}]{mcbreen06}
McBreen S., Hanlon L., McGlynn S., McBreen B., Foley S., Preece R., von Kienlin A., Williams O.~R., 2006, \aap, 455, 433

\bibitem[{McGlynn} {et al.}(2007)]{mcglynn07}
McGlynn S. et al., 2007, \aap, 466, 895

\bibitem[{McGlynn} {et al.}(2009)]{mcglynn09}McGlynn S. et al., 2009, \aap, 499, 465

\bibitem[{{Mereghetti} {et~al.}(2003){Mereghetti} {G{\"o}tz} {Borkowski}
	{Walter} {Pedersen}}]{ibas}
{Mereghetti} S., {G{\"o}tz} D., {Borkowski} J., 
	{Walter} R., {Pedersen} H., 2003, \aap,  411, L291

\bibitem[{{Mereghetti} {et~al.}(2006)}]{mereghetti06}
Mereghetti S., Paizis A., G\"otz D., Petry D., Mowlavi N., Beck M., Borkowski J., 2006, GCN, 5834

\bibitem[{{Myers} {\& Pospelov}(2003)}]{myers03}
Myers R.C., Pospelov M., 2003, Phys. Rev. Lett., 90, 211601

\bibitem[{{Nakar} {et al.}(2003)}]{nakar03}
Nakar E., Piran T., Waxman E., 2003, JCAP, 10, 005

	
%


%
%
%


\bibitem[{{Rees} {\& M{\'e}sz{\'a}ros}(1994)}]{rees94}
Rees M.J., M{\'e}sz{\'a}ros P., 1994, \apjl, 430, L93


\bibitem[{{Rhoads} {}(1997){Rhoads}}]{rhoads97}
{Rhoads} J.~E., 1997, \apjl, 487, L1


\bibitem[{{Savaglio} {et al.}(2009){Savaglio}, S. and {Glazebrook}, K. and {Le Borgne}, D.}]{savaglio09}
{Savaglio} S., {Glazebrook} K., {Le Borgne} D., 2009, \apj, 691, 182

\bibitem[{{Schlafly} {\& Finkbeiner}(2011)}]{schlafly11}
Schlafly E.~F., Finkbeiner D.~P., 2011, \apj, 737, 103

\bibitem[{{Spruit} {et al.}(2001)}]{spruit01}Spruit H.C., Daigne F., Drenkhahn G., 2001, \aap, 369, 694


%
%




\bibitem[{{Toma} {et~al.} (2012)}]{toma12}
Toma K. et al., 2012, \prd, 109, 24, 241104

\bibitem[{{Ubertini} {et~al.}(2003){Ubertini}, {Lebrun}, {Di Cocco}, {Bazzano},
  {Bird}, {Broenstad}, {Goldwurm}, {La Rosa}, {Labanti}, {Laurent}, {Mirabel},
  {Quadrini}, {Ramsey}, {Reglero}, {Sabau}, {Sacco}, {Staubert}, {Vigroux},
  {Weisskopf}, \& {Zdziarski}}]{ibis}
{Ubertini} P. {et~al.}, 2003, \aap, 411, L131






\bibitem[{{Vianello} {et al.}(2009)}]{vianello09}
Vianello G., G\"otz D., Mereghetti S., 2009, \aap, 495, 1005



\bibitem[{{Yonetoku} {et al.}(2010)}]{yonetoku10}
Yonetoku D. et al., 2010, PASJ, 62, 1495

\bibitem[{{Yonetoku} {et al.}(2011)}]{yonetoku11}
{Yonetoku} D. et al., 2011, \apj, 743, L30

\bibitem[{{Yonetoku} {et al.}(2012)}]{yonetoku12}
{Yonetoku} D. et al., 2012, \apj, 758, L1

%

%

\bibitem[{{Winkler} {et~al.}(2003){Winkler}, {Courvoisier}, {Di Cocco},
  {Gehrels}, {Gim{\' e}nez}, {Grebenev}, {Hermsen}, {Mas-Hesse}, {Lebrun},
  {Lund}, {Palumbo}, {Paul}, {Roques}, {Schnopper}, {Sch{\" o}nfelder},
  {Sunyaev}, {Teegarden}, {Ubertini}, {Vedrenne}, \& {Dean}}]{integral}
{Winkler} C. {et~al.}, 2003, \aap, 411, L1

\bibitem[{{Woosley} {et~al.}(1993){Woosley}, S.~E. and {Langer}, N. and {Weaver}, T.~A.}]{woosley93}
Woosley S.~E., Langer N., Weaver T.~A., 1993, \apj, 411, 823

\bibitem[{{Zhang} {et al.}(2011)}]{zhang11}
Zhang B.-B. et al., 2011, \apj, 730, 141


\end{thebibliography}



\label{lastpage}

\end{document}